\documentclass[aps,showpacs,preprintnumbers,amsmath,amssymb,nofootinbib]{revtex4}
\usepackage{graphicx}

\begin{document}

\preprint{gr-qc/0401049}

\title{
Vacuum solutions of the gravitational field equations in the brane
world model}

\author{T. Harko}
\email{harko@hkucc.hku.hk} \affiliation{ Department of Physics,
The University of Hong Kong, Pokfulam Road, Hong Kong}

\author{M. K. Mak}
\email{mkmak@vtc.edu.hk} \affiliation{ Department of Physics, The
University of Hong Kong, Pokfulam Road, Hong Kong}

\date{January 13, 2004}

%% REVTEX4

\begin{abstract}

We consider some classes of solutions of the static, spherically
symmetric gravitational field equations in the vacuum in the brane
world scenario, in which our Universe is a three-brane embedded in
a higher dimensional space-time. The vacuum field equations on the
brane are reduced to a system of two ordinary differential
equations, which describe all the geometric properties of the
vacuum as functions of the dark pressure and dark radiation terms
(the projections of the Weyl curvature of the bulk, generating
non-local brane stresses). Several classes of exact solutions of
the vacuum gravitational field equations on the brane are derived.
In the particular case of a vanishing dark pressure the
integration of the field equations can be reduced to the
integration of an Abel type equation. A perturbative procedure,
based on the iterative solution of an integral equation, is also
developed for this case. Brane vacuums with particular symmetries
are investigated by using Lie group techniques. In the case of a
static vacuum  brane admitting a one-parameter group of conformal
motions the exact solution of the field equations can be found,
with the functional form of the dark radiation and pressure terms
uniquely fixed by the symmetry. The requirement of the invariance
of the field equations with respect to the quasi-homologous group
of transformations also imposes a unique, linear proportionality
relation between the dark energy and dark pressure. A homology
theorem for the static, spherically symmetric gravitational field
equations in the vacuum on the brane is also proven.

\end{abstract}

%% REVTEX4
\pacs{04.50.+h, 04.20.Jb, 04.20.Cv}

\maketitle

\section{Introduction}

The idea, proposed in \cite{RS99a},  that our four-dimensional
Universe might be a three-brane, embedded in a five-dimensional
space-time (the bulk), has attracted a considerable interest in
the past few years. According to the brane-world scenario, the
physical fields (electromagnetic, Yang-Mills etc.) in our
four-dimensional Universe are confined to the three brane. These
fields are assumed to arise as fluctuations of branes in string
theories. Only gravity can freely propagate in both the brane and
bulk space-times, with the gravitational self-couplings not
significantly modified. This model originated from the study of a
single $3$-brane embedded in five dimensions, with the $5D$ metric
given by $ds^{2}=e^{-f(y)}\eta _{\mu \nu }dx^{\mu }dx^{\nu
}+dy^{2}$, which, due to the appearance of the warp factor, could
produce a large hierarchy between the scale of particle physics
and gravity. Even if the fifth dimension is uncompactified,
standard $4D$ gravity is reproduced on the brane. Hence this model
allows the presence of large, or even infinite non-compact extra
dimensions. Our brane is identified to a domain wall in a
$5$-dimensional anti-de Sitter space-time.

The Randall-Sundrum model was inspired by superstring theory. The
ten-dimensional $E_{8}\times E_{8}$ heterotic string theory, which contains
the standard model of elementary particle, could be a promising candidate
for the description of the real Universe. This theory is connected with an
eleven-dimensional theory, the $M$-theory, compactified on the orbifold $%
R^{10}\times S^{1}/Z_{2}$ \cite{HW96}. In this model we have two separated
ten-dimensional manifolds. For a review of dynamics and geometry of brane
Universes see \cite{Ma01}.

Due to the correction terms coming from the extra dimensions, significant
deviations from the Einstein theory occur in brane world models at very high
energies \cite{SMS00}, \cite{SSM00}. Gravity is largely modified at the
electro-weak scale $1$ TeV. The cosmological implications of the brane world
theories have been extensively investigated in the physical literature \cite
{all}. Gravitational collapse can also produce high energies, with the five
dimensional effects playing an important role in the formation of black
holes \cite{all1}.

For standard general relativistic spherical compact objects the exterior
space-time is described by the Schwarzschild metric. In the five dimensional
brane world models, the high energy corrections to the energy density,
together with the Weyl stresses from bulk gravitons, imply that on the brane
the exterior metric of a static star is no longer the Schwarzschild metric
\cite{Da00}. The presence of the Weyl stresses also means that the matching
conditions do not have a unique solution on the brane; the knowledge of the
five-dimensional Weyl tensor is needed as a minimum condition for
uniqueness. Static, spherically symmetric exterior vacuum solutions of the
brane world models have been proposed first by Dadhich et al. \cite{Da00}
and Germani and Maartens \cite{GeMa01}. The first of these solutions,
obtained in \cite{Da00}, has the mathematical form of the Reissner-Nordstrom
solution, in which a tidal Weyl parameter plays the role of the electric
charge of the general relativistic solution. The solution has been obtained
by imposing the null energy condition on the 3-brane for a bulk having non
zero Weyl curvature. The solution can be matched to the interior solution
corresponding to a constant density brane world star. A second exterior
solution, which also matches a constant density interior, has been derived
in \cite{GeMa01}.

Two families of analytic solutions of the spherically symmetric vacuum brane
world model equations (with $g_{tt}\neq -1/g_{rr}$), parameterized by the
ADM mass and a PPN parameter $\beta $ have been obtained by Casadio, Fabri
and Mazzacurati \cite{Ca02}. Non-singular black-hole solutions in the brane
world model have been considered in \cite{Da03}, by relaxing the condition
of the zero scalar curvature but retaining the null energy condition. The
``on brane'' 4-dimensional Gauss and Codazzi equations for an arbitrary
static spherically symmetric star in a Randall--Sundrum type II brane world
have been completely solved by Visser and Wiltshire \cite{Vi03}. The
on-brane boundary can be used to determine the full $5$-dimensional
space-time geometry. The procedure can be generalized to solid objects such
as planets. A method to extend into the bulk asymptotically flat static
spherically symmetric brane-world metrics has been proposed by Casadio and
Mazzacurati \cite{Ca03}. The exact integration of the field equations along
the fifth coordinate was done by using the multipole ($1/r$) expansion. The
results show that the shape of the horizon of the brane black hole solutions
is very likely a flat ``pancake'' for astrophysical sources.

Stellar structure in brane worlds is very different from that in ordinary
general relativity. An exact interior uniform-density stellar solution on
the brane has been found in \cite{GeMa01}. In this model the general
relativistic upper bound for the mass-radius ratio, $M/R<4/9$, is reduced by
5-dimensional high-energy effects. The existence of brane world neutron
stars leads to a constraint on the brane tension, which is stronger than the
big-bang nucleosynthesis constraint, but weaker than the Newton-law
experimental constraints \cite{GeMa01}.

It is the purpose of the present paper to systematically consider
spherically symmetric space-times in vacuum on the brane. As a first step we
derive the two basic ordinary differential equations for the dark radiation
and dark pressure, describing the geometry of the vacuum on the brane. By
means of some appropriate transformations these equations take the form of
an autonomous system of two ordinary differential equations. Some simple
integrability cases are considered, leading to some \ already known or new
vacuum solutions on the brane. The very important case corresponding to a
vanishing dark pressure term is considered in detail. The integration of the
gravitational field equations in the vacuum on the brane is reduced to the
integration of an Abel type equation. Since this equation does not satisfy
the known integrability conditions, the solution is obtained in terms of
perturbative series obtained by solving the integral equation associated to
this problem.

Next we consider vacuum space-times on the brane that are related to some
particular Lie groups of transformations. As a first group of admissible
transformations for the vacuum on the brane we shall consider spherically
symmetric and static solutions of the gravitational field equations that
admits a one-parameter group of conformal motions, i.e., the metric tensor $%
g_{\mu \nu }$ has the property $L_{\xi }g_{\mu \nu }=\phi \left( r\right)
g_{\mu \nu }$, where the left-hand side is the Lie derivative of the metric
tensor, describing the gravitational field on the vacuum brane, with respect
to the vector field $\xi ^{\mu }$, and $\phi $ is an arbitrary function of
the radial coordinate $r$. With this assumption the gravitational field
equations describing the static vacuum brane can be integrated in
Schwarzschild coordinates, and an exact simple solution, corresponding to a
brane admitting a one-parameter group of motions can be obtained.

Suppose that from some static, spherically symmetric solution of the vacuum
gravitational field equations on the brane we have obtained we want to
construct other physical solutions of the field equations by means of scale
transformations. The process of constructing a new physical model by
applying scale changes to the given initial model is referred to as a
''homology transformation'' \cite{Cha59}. The homology properties of stars
have been intensively investigated in astrophysics and families of stars
constructed in such a way from a given star are called homologous stars. For
Newtonian homologous stars in equilibrium the individual members are related
to each other by transformations of the form $r\rightarrow \bar{r}=ar$, $%
\rho \rightarrow \bar{\rho}=b\rho $ and $M\rightarrow \bar{M}=cM$, with $%
a,b,c$ constants. Chandrasekhar \cite{Cha59} refers to this change of scale
as a ''homologous transformation'', and the homology theorem of
Chandrasekhar \cite{Cha59} states that if $\theta \left( \xi \right) $ is a
solution of the stellar structure equations then so is $C^{2/\left(
n-1\right) }\theta \left( C\xi \right) $ also, where $C$ is an arbitrary
constant and $1<n\leq 5$. By analyzing the homology transformation
properties of the gravitational field equations on the static vacuum brane,
by using Lie group theory techniques, we shall show that the requirement of
the invariance of the field equations with respect to an infinitesimal
generator ${\bf X}$ fixes in an unique way the relation between the dark
pressure and the dark radiation terms. We also prove the homology theorem
for the gravitational field equations in vacuum on the brane.

The present paper is organized as follows. The basic equations describing
the spherically symmetric gravitational field equations in the vacuum on the
brane are derived in Section II. Some particular classes of solutions for
vacuum branes are obtained in Section III. In Section IV we consider vacuum
brane space-times admitting a one parameter group of conformal motions.
Homology properties of the gravitational field equations are investigated in
Section V. We conclude and discuss our results in Section VI.

\section{Static, spherically symmetric vacuum field
equations on the brane}

On the $5$-dimensional space-time (the bulk), with the negative vacuum
energy $\Lambda _{5}$ and brane energy-momentum as source of the
gravitational field, the Einstein field equations are given by
\begin{equation}
G_{IJ}=k_{5}^{2}T_{IJ},\qquad T_{IJ}=-\Lambda _{5}g_{IJ}+\delta (Y)\left[
-\lambda _{b}g_{IJ}+T_{IJ}^{\text{matter}}\right] ,
\end{equation}
with $\lambda _{b}$ the vacuum energy on the brane and $k_{5}^{2}=8\pi G_{5}$%
. In this space-time a brane is a fixed point of the $Z_{2}$ symmetry. In
the following capital Latin indices run in the range $0,...,4$, while Greek
indices take the values $0,...,3$.

Assuming a metric of the form $ds^{2}=(n_{I}n_{J}+g_{IJ})dx^{I}dx^{J}$, with
$n_{I}dx^{I}=d\chi $ the unit normal to the $\chi =$constant hypersurfaces
and $g_{IJ}$ the induced metric on $\chi =$constant hypersurfaces, the
effective four-dimensional gravitational equations on the brane (the Gauss
equation), take the form \cite{SMS00,SSM00}:
\begin{equation}
G_{\mu \nu }=-\Lambda g_{\mu \nu }+k_{4}^{2}T_{\mu \nu }+k_{5}^{4}S_{\mu \nu
}-E_{\mu \nu },  \label{Ein}
\end{equation}
where $S_{\mu \nu }$ is the local quadratic energy-momentum correction
\begin{equation}
S_{\mu \nu }=\frac{1}{12}TT_{\mu \nu }-\frac{1}{4}T_{\mu }{}^{\alpha }T_{\nu
\alpha }+\frac{1}{24}g_{\mu \nu }\left( 3T^{\alpha \beta }T_{\alpha \beta
}-T^{2}\right) ,
\end{equation}
and $E_{\mu \nu }$ is the non-local effect from the free bulk gravitational
field, the transmitted projection of the bulk Weyl tensor $C_{IAJB}$, $%
E_{IJ}=C_{IAJB}n^{A}n^{B}$, with the property $E_{IJ}\rightarrow E_{\mu \nu
}\delta _{I}^{\mu }\delta _{J}^{\nu }\quad $as$\quad \chi \rightarrow 0$. We
have also denoted $k_{4}^{2}=8\pi G$, with $G$ the usual four-dimensional
gravitational constant.

The four-dimensional cosmological constant, $\Lambda $, and the
four-dimensional coupling constant, $k_{4}$, are given by $\Lambda
=k_{5}^{2}\left( \Lambda _{5}+k_{5}^{2}\lambda _{b}^{2}/6\right) /2$ and $%
k_{4}^{2}=k_{5}^{4}\lambda _{b}/6$, respectively. In the limit $\lambda
_{b}^{-1}\rightarrow 0$ we recover standard general relativity.

The Einstein equation in the bulk and the Codazzi equation also imply the
conservation of the energy-momentum tensor of the matter on the brane, $%
D_{\nu }T_{\mu }{}^{\nu }=0$, where $D_{\nu }$ denotes the brane covariant
derivative. Moreover, from the contracted Bianchi identities on the brane it
follows that the projected Weyl tensor should obey the constraint $D_{\nu
}E_{\mu }{}^{\nu }=k_{5}^{4}D_{\nu }S_{\mu }{}^{\nu }$.

The symmetry properties of $E_{\mu \nu }$ imply that in general we can
decompose it irreducibly with respect to a chosen $4$-velocity field $u^{\mu
}$ as \cite{Ma01}
\begin{equation}
E_{\mu \nu }=-k^{4}\left[ U\left( u_{\mu }u_{\nu }+\frac{1}{3}h_{\mu \nu
}\right) +P_{\mu \nu }+2Q_{(\mu }u_{\nu )}\right] ,  \label{WT}
\end{equation}
where $k=k_{5}/k_{4}$, $h_{\mu \nu }=g_{\mu \nu }+u_{\mu }u_{\nu }$ projects
orthogonal to $u^{\mu }$, the ''dark radiation'' term $U=-k^{4}E_{\mu \nu
}u^{\mu }u^{\nu }$ is a scalar, $Q_{\mu }=k^{4}h_{\mu }^{\alpha }E_{\alpha
\beta }$ a spatial vector and $P_{\mu \nu }=-k^{4}\left[ h_{(\mu }\text{ }%
^{\alpha }h_{\nu )}\text{ }^{\beta }-\frac{1}{3}h_{\mu \nu }h^{\alpha \beta }%
\right] E_{\alpha \beta }$ a spatial, symmetric and trace-free tensor.

In the case of the vacuum state with $\rho =p=0$, $T_{\mu \nu }\equiv 0$ and
consequently $S_{\mu \nu }=0$. Therefore, by neglecting the effect of the
cosmological constant, the field equations describing a static brane take
the form
\begin{equation}
R_{\mu \nu }=-E_{\mu \nu },
\end{equation}
with $R_{\mu }^{\mu }=0=E_{\mu }^{\mu }$. In the vacuum case $E_{\mu \nu }$
satisfies the constraint $D_{\nu }E_{\mu }{}^{\nu }=0$. In an inertial frame
at any point on the brane we have $u^{\mu }=\delta _{0}^{\mu }$ and $h_{\mu
\nu }=$diag$\left( 0,1,1,1\right) $. In a static vacuum $Q_{\mu }=0$ and the
constraint for $E_{\mu \nu }$ takes the form
\begin{equation}
\frac{1}{3}D_{\mu }U+\frac{4}{3}UA_{\mu }+D^{\nu }P_{\mu \nu }+A^{\nu
}P_{\mu \nu }=0,
\end{equation}
where $D_{\mu }$ is the projection (orthogonal to $u^{\mu }$) of the
covariant derivative and $A_{\mu }=u^{\nu }D_{\nu }u_{\mu }$ is the
4-acceleration. In the static spherically symmetric case we may choose $%
A_{\mu }=A(r)r_{\mu }$ and $P_{\mu \nu }=P(r)\left( r_{\mu }r_{\nu }-\frac{1%
}{3}h_{\mu \nu }\right) $, where $A(r)$ and $P(r)$ (the ''dark pressure'')
are some scalar functions of the radial distance $r$, and $r_{\mu }$ is a
unit radial vector.

We chose the static spherically symmetric metric on the brane in the form
\begin{equation}
ds^{2}=-e^{\nu \left( r\right) }dt^{2}+e^{\lambda \left( r\right)
}dr^{2}+r^{2}\left( d\theta ^{2}+\sin ^{2}\theta d\phi ^{2}\right) .
\label{line}
\end{equation}

Then the gravitational field equations and the effective energy-momentum
tensor conservation equation in the vacuum take the form
\begin{equation}
-e^{-\lambda }\left( \frac{1}{r^{2}}-\frac{\lambda ^{\prime }}{r}\right) +%
\frac{1}{r^{2}}=\frac{48\pi G}{k^{4}\lambda _b}U,  \label{f1}
\end{equation}
\begin{equation}
e^{-\lambda }\left( \frac{\nu ^{\prime }}{r}+\frac{1}{r^{2}}\right) -\frac{1%
}{r^{2}}=\frac{16\pi G}{k^{4}\lambda _b}\left( U+2P\right) ,
\label{f2}
\end{equation}
\begin{equation}
e^{-\lambda }\left( \nu ^{\prime \prime }+\frac{\nu ^{\prime 2}}{2}+\frac{%
\nu ^{\prime }-\lambda ^{\prime }}{r}-\frac{\nu ^{\prime }\lambda ^{\prime }%
}{2}\right) =\frac{32\pi G}{k^{4}\lambda _b}\left( U-P\right) ,
\label{f3}
\end{equation}
\begin{equation}
\nu ^{\prime }=-\frac{U^{\prime }+2P^{\prime }}{2U+P}-\frac{6P}{r\left(
2U+P\right) },  \label{f4}
\end{equation}
where $^{\prime }=d/dr$. Eq. (\ref{f1}) can immediately be integrated to give
\begin{equation}
e^{-\lambda }=1-\frac{C_{1}}{r}-\frac{Q\left( r\right) }{r},  \label{m1}
\end{equation}
where $C_{1}$ is an arbitrary constant of integration, and we denoted
\begin{equation}
Q\left( r\right) =\frac{48\pi G}{k^{4}\lambda _b}\int r^{2}U\left(
r\right) dr.
\end{equation}

The function $Q$ is the gravitational mass corresponding to the
dark radiation term (the dark mass). For $U=0$ the metric
coefficient given by Eq. (\ref{m1}) must tend to the standard
general relativistic Schwarzschild metric coefficient, which gives
$C_{1}=2GM$, where $M=$constant is the mass of the gravitating
body. In the following we also denote $\alpha =16\pi
G/k^{4}\lambda _b$. By substituting $\nu ^{\prime }$ given by Eq.
(\ref{f4}) into Eq. (\ref{f2}) and with the use of Eq. (\ref{m1})
we obtain the following system of differential equations satisfied
by the dark radiation term $U$, the dark pressure $P$ and the dark
mass $Q$, describing the vacuum gravitational field, exterior to a
massive body, in the brane world model:
\begin{equation}
\frac{dU}{dr}=-\frac{\left( 2U+P\right) \left[ 2GM+Q+\alpha \left(
U+2P\right) r^{3}\right] }{r^{2}\left( 1-\frac{2GM}{r}-\frac{Q}{r}\right) }-2%
\frac{dP}{dr}-\frac{6P}{r},  \label{e1}
\end{equation}
\begin{equation}
\frac{dQ}{dr}=3\alpha r^{2}U.  \label{e2}
\end{equation}

The system of equations (\ref{e1}) and (\ref{e2}) can be transformed to an
autonomous system of differential equations by means of the transformations
\begin{equation}
q=\frac{2GM+Q}{r},\mu =3\alpha r^{2}U,p=3\alpha r^{2}P,\theta =\ln r.
\label{trans}
\end{equation}

With the use of the new variables given by Eqs. (\ref{trans}), Eqs. (\ref{e1}%
) and (\ref{e2}) become
\begin{equation}
\frac{dq}{d\theta }=\mu -q,  \label{aut1}
\end{equation}
\begin{equation}
\frac{d\mu }{d\theta }=-\frac{\left( 2\mu +p\right) \left[ q+\frac{1}{3}%
\left( \mu +2p\right) \right] }{1-q}-2\frac{dp}{d\theta }+2\mu -2p.
\label{aut2}
\end{equation}

Eqs. (\ref{e1}) and (\ref{e2}), or, equivalently, (\ref{aut1}) and
(\ref{aut2}) may be called the structure equations of the vacuum
on the brane.

\section{Classes of vacuum solutions on the brane}

The system of structure equations (\ref{e1})-(\ref{e2}) is not
closed until a further condition is imposed on the functions $U$
and $P$. By choosing some particular forms of these functions
several classes of static vacuum solutions can be generated in the
framework of the brane world model. As a first case we consider
that the dark radiation $U$ and the dark pressure $P$ satisfy the
constraint
\begin{equation}
2U+P=0.
\end{equation}

Then Eq. (\ref{e1}) takes the form
\begin{equation}
\frac{dP}{dr}=-4\frac{P}{r},
\end{equation}
with the general solution given by
\begin{equation}
P=\frac{P_{0}}{r^{4}},U=-\frac{P_{0}}{2r^{4}}
\end{equation}
where $P_{0}$ is an arbitrary constant of integration. Eq.
(\ref{e2}) immediately gives the dark mass as
\begin{equation}
Q=\frac{3\alpha P_{0}}{2r}+Q_{0},
\end{equation}
where $Q_{0}$ is a constant of integration. Therefore the metric on the
brane is
\begin{equation}
e^{\nu }=e^{-\lambda }=1-\frac{2GM+Q_{0}}{r}-\frac{3\alpha P_{0}}{2r^{2}}.
\end{equation}

This form of the metric has been first obtained by Dadhich et al \cite{Da00}.

A second class of solutions of the system of equations (\ref{e1})-(\ref{e2})
can be obtained by assuming that
\begin{equation}
U+2P=0.
\end{equation}

Then Eq. (\ref{e1}) is transformed into an algebraic equation, which gives
\begin{equation}
Q=\frac{2}{3}r-2GM.
\end{equation}

Hence the metric coefficients of the vacuum brane line element becomes
\begin{equation}
e^{\nu }=C_{0}r^{2},e^{-\lambda }=\frac{1}{3},  \label{solp1}
\end{equation}
where $C_{0}$ is a constant of integration and $e^{\nu }$ has been obtained
by integrating Eq. (\ref{f4}). For this class of solutions the projections
of the Weyl bulk tensor are given by
\begin{equation}
U=-2P=\frac{2}{9\alpha r^{2}}.
\end{equation}

In the case of a vanishing dark radiation, $U=0$, which also implies a
vanishing dark mass $Q=0$, the dark pressure $P$ satisfies a Bernoulli type
equation, given by
\begin{equation}
\frac{dP}{dr}+\frac{3P}{r}+\frac{P\left( GM+\alpha r^{3}P\right) }{%
r^{2}\left( 1-\frac{2GM}{r}\right) }=0,
\end{equation}
with the general solution
\begin{equation}
P=\frac{1}{r^{3}\left( C_{1}\sqrt{1-\frac{2GM}{r}}-\frac{\alpha }{GM}\right)
},
\end{equation}
where $C_{1}$ is an arbitrary constant of integration. Eq. (\ref{f4}) gives $%
\nu ^{\prime }=-2P^{\prime }/P-6/r$, or $\exp \left( \nu \right)
=C_{2}/P^{2}r^{6}$, where $C_2$ is an arbitrary constant of
integration. Hence for $U=0$ the metric tensor components are
given by
\begin{equation}
e^{\nu }=C_{2}\left( 1-\frac{2GM}{r}\right) \left[ C_{1}-\frac{\alpha }{GM}%
\left( 1-\frac{2GM}{r}\right) ^{-1/2}\right] ^{2},e^{-\lambda }=1-\frac{2GM}{%
r}.
\end{equation}

Since $\alpha /GM$ is very small, for a zero dark radiation, $U=0$, the
deviations from the Schwarzschild geometry are very small. The standard
general relativistic result is recovered for $\alpha \rightarrow 0$, which
gives $C_{1}^{2}C_{2}=1$.

Next we consider the case of a vanishing dark pressure $P=0$. The dark
radiation and the dark mass can be obtained by solving the following system
of coupled differential equations, which immediately follows from Eqs. (\ref
{aut1}) and (\ref{aut2}):
\begin{equation}
\frac{dq}{d\theta }=\mu -q,  \label{p01}
\end{equation}
\begin{equation}
\frac{d\mu }{d\theta }=2\mu \frac{3-6q-\mu }{3\left( 1-q\right) }.
\label{p02}
\end{equation}

By eliminating $\mu $ between Eqs. (\ref{p01}) and (\ref{p02}) we obtain the
following second order differential equation for the function $q$:
\begin{equation}\label{q1}
3\left( 1-q\right) \frac{d^{2}q}{d\theta ^{2}}+\left( 13q-3\right) \frac{dq}{%
d\theta }+2\left( \frac{dq}{d\theta }\right) ^{2}+2q\left( 7q-3\right) =0.
\end{equation}

By means of the successive transformations $dq/d\theta =1/v$, $%
v=w\left( 1-q\right) ^{-2/3}$ equation (\ref{q1}) can be transformed to the
following, Abel type, first order differential equation:
\begin{equation}
\frac{dw}{dq}-\frac{13q-3}{3}\left( 1-q\right) ^{-5/3}w^{2}+\frac{2q\left(
3-7q\right) }{3}\left( 1-q\right) ^{-7/3}w^{3}=0.  \label{q2}
\end{equation}

It is a matter of simple calculations to check that Eq. (\ref{q2}) has a
particular solution of the form:
\begin{equation}
w=-\frac{1}{q}\left( 1-q\right) ^{2/3}.  \label{qp}
\end{equation}

By introducing a new variable $\eta =\left( 1-q\right) ^{-1/3}$, $q=1-\eta
^{-3}$, Eq. (\ref{q2}) can be further transformed to
\begin{equation}
\frac{dw}{d\eta }-\frac{10\eta ^{3}-13}{\eta ^{2}}w^{2}+\frac{2\left( \eta
^{3}-1\right) \left( 7-4\eta ^{3}\right) }{\eta ^{3}}w^{3}=0.  \label{w}
\end{equation}
$\allowbreak $

Therefore we have reduced the problem of the integration of the
gravitational field equations for the vacuum on the brane in the case of a
vanishing dark pressure term, $P=0$, to the problem of the integration of an
Abel type equation. However, Eq. (\ref{w}) does not satisfy the standard
integration conditions for Abel type equations \cite{Makm}, and an exact
analytical solution of this equation seems to be difficult to obtain. Hence,
in order to find some explicit solutions for the vacuum \ gravitational
field on the brane we have to use some perturbative methods.

By using the Laplace transform and the convolution theorem, the differential
equation (\ref{q1}) is equivalent to the following integral equation,
\begin{equation}
q\left( \theta \right) =\int_{\theta _{0}}^{\theta }F\left( \theta
-x\right) \left[ 3q\frac{d^{2}q}{dx^{2}}-13q\frac{dq}{dx}-2\left(
\frac{dq}{dx}\right) ^{2}-14q^{2}\right] dx+q_{0}\left( \theta
\right) ,  \label{int}
\end{equation}
where
\begin{equation}
F\left( \theta -x\right) =\frac{1}{9}\left[ e^{2\left( \theta -x\right)
}-e^{-\left( \theta -x\right) }\right] ,
\end{equation}
\begin{equation}
q_{0}\left( \theta \right) =A_{1}e^{-\theta }+A_{2}e^{2\theta },
\end{equation}
and we denoted $A_{1}=\left[ 3q\left( \theta _{0}\right) -\mu \left( \theta
_{0}\right) \right] \exp \left( \theta _{0}\right) /3$ and $A_{2}=\mu \left(
\theta _{0}\right) \exp \left( -2\theta _{0}\right) /3$. $\theta _{0}=\ln
r_{0}$ is an arbitrary point, like, for example, the vacuum boundary of a
compact astrophysical object, in which the functions $q(r)$ and $\mu \left(
r\right) $ take the values $q\left( r_{0}\right) =\left[ 2GM+Q\left(
r_{0}\right) \right] /r_{0}$ and $\mu \left( r_{0}\right) =3\alpha
r_{0}^{2}U\left( r_{0}\right) $, respectively.

The solution of the integral equation (\ref{int}) can be easily obtained by
using the method of successive approximations, or the method of iterations.
In this way we can generate the solution to any desired degree of accuracy.
Taking as an initial approximation the general solution of the linear part
of the differential equation (\ref{q1}), the general solution of the
integral equation (\ref{int}) can be expressed in the first, second and $m$%
th order approximation, $m\in N$, as follows:
\begin{equation}
q_{1}\left( \theta \right) =\int_{\theta _{0}}^{\theta }F\left( \theta
-x\right) \left[ 3q_{0}\frac{d^{2}q_{0}}{dx^{2}}-13q_{0}\frac{dq_{0}}{dx}%
-2\left( \frac{dq_{0}}{dx}\right) ^{2}-14q_{0}^{2}\right] dx+q_{0}\left(
\theta \right) ,
\end{equation}
\[
.\text{ \ }.\text{ \ }.
\]
\begin{equation}
q_{m}\left( \theta \right) =\int_{\theta _{0}}^{\theta }F\left( \theta
-x\right) \left[ 3q_{m-1}\frac{d^{2}q_{m-1}}{dx^{2}}-13q_{m-1}\frac{dq_{m-1}%
}{dx}-2\left( \frac{dq_{m-1}}{dx}\right) ^{2}-14q_{m-1}^{2}\right]
dx+q_{m-1}\left( \theta \right) ,
\end{equation}
\begin{equation}
q\left( \theta \right) =\lim_{m\rightarrow \infty }q_{m}\left( \theta
\right) .
\end{equation}

The zero'th order approximation to the solution of the static spherically
symmetric gravitational field equations in the vacuum on the brane is given
by
\begin{equation}
e^{\nu }=\frac{C_{0}}{\sqrt{U}}=C_{0}\sqrt{\frac{\alpha }{A_{2}}},
\end{equation}
\begin{equation}
e^{-\lambda }=1-\frac{A_{1}}{r}-A_{2}r^{2},
\end{equation}
\begin{equation}
U=\frac{A_{2}}{\alpha },
\end{equation}
where $C_{0}$ is an arbitrary constant of integration.

The first order approximation to the solution is given by
\begin{equation}
e^{\nu }=C_{0}\sqrt{\frac{\alpha r_{0}}{2}}\sqrt{\frac{r}{A_{2}\left(
r_{0}-r\right) \left[ A_{1}+A_{2}r_{0}^{2}r+A_{2}r_{0}r^{2}\right] }},
\end{equation}
\begin{equation}
e^{-\lambda }=1+\frac{A_{2}r_{0}^{2}\left[ \left( 4/5\right)
A_{2}r_{0}^{3}+A_{1}\right] }{r}-3A_{1}A_{2}r-2A_{2}\left( 2A_{2}r_{0}^{2}-%
\frac{A_{1}}{r_{0}}\right) r^{2}+\frac{6}{5}A_{2}^{2}r^{4},
\end{equation}
\begin{equation}
U=\frac{2A_{2}\left( r_{0}-r\right) \left[ A_{1}+A_{2}r_{0}r\left(
r_{0}+r\right) \right] }{\alpha r_{0}r}.
\end{equation}

Therefore the general solution to the static gravitational field equations
on the vacuum brane can be obtained in any order of approximation.

\section{Static vacuum branes admitting a one-parameter group of conformal
motions}

In the present Section we are going to consider a special class of static
vacuum brane solutions, which have as a group of admissible transformations
the conformal motions (homothetic transformations).

For a spherically symmetric and static vacuum on the brane the assumption of
the existence of an one-parameter group of conformal motions requires that
the condition
\begin{equation}
L_{\xi }g_{\mu \nu }=\xi _{\mu ;\nu }+\xi _{\nu ;\mu }=\phi \left( r\right)
g_{\mu \nu },  \label{conf}
\end{equation}
holds for the metric tensor components, where the left-hand side is the Lie
derivative of the metric tensor, describing the vacuum brane gravitational
field, with respect to the vector field $\xi ^{\mu }$, and $\phi \left(
r\right) $ is an arbitrary function of the radial coordinate $r$. We shall
further restrict the field $\xi ^{\mu }$ by demanding $\xi ^{\mu }u_{\mu }=0$%
. Then as a consequence of the spherical symmetry we have $\xi ^{2}=\xi
^{3}=0$. This type of symmetry has been intensively used to describe the
interior of neutral or charged general relativistic stellar-type objects
\cite{He}. With the assumption (\ref{conf}) the gravitational field
equations describing the spherically symmetric static vacuum brane can be
integrated in Schwarzschild coordinates and an exact solution can be
obtained. Moreover, the requirement of the conformal invariance of the
static brane uniquely fix the functional form of the projections of the bulk
Weyl tensor components $U(r)$ and $P(r)$.

Using the line element (\ref{line}), the equation (\ref{conf}) explicitly
reads
\begin{equation}
\xi ^{1}\nu ^{^{\prime }}=\phi ,\xi ^{0}=\bar{C}=\text{constant},\xi ^{1}=%
\frac{\phi r}{2},\lambda ^{^{\prime }}\xi ^{1}+2\frac{d\xi ^{1}}{dr}=\phi .
\label{conf1}
\end{equation}

Equations (\ref{conf1}) have the general solution given by \cite{He}
\begin{equation}
e^{\nu }=A^{2}r^{2},\phi =Ce^{-\lambda /2},\xi ^{\mu }=\bar{C}\delta
_{0}^{\mu }+\delta _{1}^{\mu }\frac{\phi r}{2},  \label{conf2}
\end{equation}
with $A$ and $C$ arbitrary constants of integration.

Hence the requirement of the existence of conformal motions imposes strong
constraints on the form of the metric tensor coefficients of the static
vacuum brane. Substituting Eqs. (\ref{conf2}) into the field equations (\ref
{f1})-(\ref{f3}) we obtain
\begin{equation}
\frac{1}{r^{2}}\left( 1-\frac{\phi ^{2}}{C^{2}}\right) -\frac{2}{C^{2}}\frac{%
\phi \phi ^{^{\prime }}}{r}=3\alpha U,  \label{c1}
\end{equation}
\begin{equation}
\frac{1}{r^{2}}\left( 1-3\frac{\phi ^{2}}{C^{2}}\right) =-\alpha \left(
U+2P\right) ,  \label{c2}
\end{equation}
\begin{equation}
\frac{1}{C^{2}}\frac{\phi ^{2}}{r^{2}}+\frac{2}{C^{2}}\frac{\phi \phi
^{^{\prime }}}{r}=\alpha \left( U-P\right) .  \label{c3}
\end{equation}

We can formally solve the field equations (\ref{c2}) and (\ref{c3}) to
express $U$ and $P$ as
\begin{equation}
U=\frac{1}{3\alpha }\left[ \frac{4}{C^{2}}\frac{\phi \phi ^{\prime }}{r}-%
\frac{1}{r^{2}}\left( 1-\frac{5}{C^{2}}\phi ^{2}\right) \right] ,  \label{u}
\end{equation}
\begin{equation}
P=-\frac{1}{3\alpha }\left[ \frac{2}{C^{2}}\frac{\phi \phi ^{\prime }}{r}+%
\frac{1}{r^{2}}\left( 1-\frac{2}{C^{2}}\phi ^{2}\right) \right] .
\end{equation}

With the use of the Eqs. (\ref{c1}) and (\ref{u}) it follows that the
function $\phi $ satisfies the first order differential equation
\begin{equation}
\frac{3}{C^{2}}\phi \phi ^{\prime }=\frac{1}{r}\left( 1-\frac{3}{C^{2}}\phi
^{2}\right) ,
\end{equation}
with the general solution given by
\begin{equation}
\phi ^{2}=\frac{C^{2}}{3}\left( 1+\frac{B}{r^{2}}\right) ,
\end{equation}
where $B>0$ is a constant of integration. Therefore the general
solution of the static gravitational field equations on the brane
for space-times admitting an one-parameter group of conformal
motions is given by
\begin{equation}
e^{\nu }=A^{2}r^{2},e^{-\lambda }=\frac{1}{3}\left( 1+\frac{B}{r^{2}}\right)
,
\end{equation}
\begin{equation}
U=\frac{1}{9\alpha r^{2}}\left( 2+\frac{B}{r^{2}}\right) ,
\end{equation}
\begin{equation}
P=\frac{1}{9\alpha r^{2}}\left( \frac{4B}{r^{2}}-1\right) .
\end{equation}

In the case $B=0$ we recover the solution given by Eqs. (\ref{solp1}),
satisfying the condition $U+2P=0$. For this case the function $\phi =C/\sqrt{%
3}$ is a constant.

\section{Homology properties of the static gravitational field equations in
vacuum on the brane}

Let us assume that a solution of the field equations (\ref{f1})-(\ref{f4})
is known. Then it seems reasonable to require that a family of solutions
should exist, whose individual members are related by more general
transformations of the form $r\rightarrow \bar{r}\left( r\right) $, $%
U\rightarrow \bar{U}\left( U\right) $, $P\rightarrow \bar{P}\left(
P\right) $ and $Q\rightarrow $ $\bar{Q}\left( Q\right) $
\cite{Co77}. We shall call a set of solutions of the vacuum
gravitational field equations on the brane related by
transformations of this form a homologous family of solutions.

In order to obtain the homology properties of the structure equations \ Eqs. (\ref{e1}) and (\ref{e2}%
), it is necessary first to close the system of equations. We
shall do this by assuming that the dark pressure $P$ and the dark
radiation $U$ terms are related by an arbitrary functional
relation $P=P\left( U\right) $. Then, by denoting $\gamma \left(
U\right) =P\left( U\right) /U$ and $dP/dU=P^{\prime }(U)=c_{s}$,
the basic equations describing the vacuum gravitational field on
the brane take the form
\begin{equation}
\frac{dU}{dr}=-\frac{\gamma \left( U\right)
U}{1+2c_{s}}\frac{\left[ 1+2\gamma ^{-1}\left( U\right) \right]
\left\{ 2GM+Q+\alpha r^{3}\left[
1+2\gamma \left( U\right) \right] U\right\} +6r-6\left( 2GM+Q\right) }{%
r^{2}\left( 1-\frac{2GM}{r}-\frac{Q}{r}\right) },  \label{h1}
\end{equation}
\begin{equation}
\frac{dQ}{dr}=3\alpha r^{2}U.  \label{h2}
\end{equation}

A system of ordinary differential equations
\begin{equation}
\frac{dy^{k}}{dx}=f^{k}\left( x,y\right) ,k=1,2,...,m,
\end{equation}
with $y=\left( y^{1},y^{2},...,y^{m}\right) $ is invariant under the action
of the infinitesimal generator ${\bf X}=\zeta \left( x,y\right) \frac{%
\partial }{\partial x}+\eta ^{k}\left( x,y\right) \frac{\partial }{\partial
y^{k}}$ if and only if $\left[ {\bf L},{\bf X}\right] =r{\bf X}$, where $%
\left[ {}\right] $ denotes the Lie bracket, ${\bf L}=\frac{\partial }{%
\partial x}+f^{k}\frac{\partial }{\partial y^{k}}$ and $r={\bf L}\left(
\zeta \right) $ \cite{Ol93}, or, in explicit form \cite{Co77}
\begin{equation}
\frac{\partial \eta ^{k}}{\partial x}+f^{j}\frac{\partial \eta ^{k}}{%
\partial y^{j}}-f^{k}\frac{\partial \zeta }{\partial x}-f^{k}f^{j}\frac{%
\partial \zeta }{\partial y^{j}}=\zeta \frac{\partial f^{k}}{\partial x}%
+\eta ^{j}\frac{\partial f^{k}}{\partial y^{j}},k=1,2,...,m.  \label{2.1}
\end{equation}

In the particular case where ${\bf X}$ generates quasi-homologous
transformations of the form $x\rightarrow \bar{x}\left( x\right) $, $%
y^{j}\rightarrow \bar{y}^{j}\left( y^{j}\right) $we have $\zeta =\zeta
\left( x\right) $ and $\eta ^{j}=\eta ^{j}\left( y^{j}\right) $. As a result
Eq. (\ref{2.1}) becomes
\begin{equation}
\frac{d\eta ^{k}\left( y^{k}\right) }{dy^{k}}-\frac{d\zeta \left( x\right) }{%
dx}={\bf X}\left( \ln \left| f^{k}\right| \right) ,  \label{mod}
\end{equation}
with no sum over $k$.

To analyze the homologous behavior of the Eqs. (\ref{h1}) and (\ref{h2})
with respect to quasi-homologous transformations involving a general
functional dependence of the physical parameters, $r=\bar{r}\left( r\right) $%
, $U=\bar{U}\left( U\right) $, $P=\bar{P}\left( P\right) $ and $Q=\bar{Q}%
\left( Q\right) $, we shall investigate the group of transformations
generated by the infinitesimal generator
\begin{equation}
{\bf X}=\zeta \left( r\right) \frac{\partial }{\partial r}+\eta ^{1}\left(
U\right) \frac{\partial }{\partial U}+\eta ^{2}\left( Q\right) \frac{%
\partial }{\partial Q}.  \label{gen}
\end{equation}

As applied to the case of Eqs. (\ref{h1}) and (\ref{h2}), Eqs. (\ref{mod})
give
\begin{eqnarray}\label{X1}
\frac{d\eta ^{1}\left( U\right) }{dU}-\frac{d\zeta }{dr} &=&\zeta
\left(
r\right) \frac{\partial }{\partial r}\ln \left\{ \frac{\left[ 1+2\gamma ^{-1}%
\right] \left[ 2GM+Q+\alpha r^{3}\left( 1+2\gamma \right) U\right]
+6r-6\left( 2GM+Q\right) }{r^{2}\left( 1-\frac{2GM}{r}-\frac{Q}{r}\right) }%
\right\} + \nonumber\\
&&\eta ^{1}\left( U\right) \frac{\partial }{\partial U}\ln \left\{ \gamma U%
\frac{\left[ 1+2\gamma ^{-1}\right] \left[ 2GM+Q+\alpha
r^{3}\left( 1+2\gamma
\right) U\right] +6r-6\left( 2GM+Q\right) }{1+2c_{s}}\right\} +  \nonumber \\
&&\eta ^{2}\left( Q\right) \frac{\partial }{\partial Q}\ln \left\{ \frac{%
\left[ 1+2\gamma ^{-1}\right] \left[ 2GM+Q+\alpha r^{3}\left(
1+2\gamma
\right) U\right] +6r-6\left( 2GM+Q\right) }{\left( 1-\frac{2GM}{r}-\frac{Q}{r}%
\right) }\right\} ,
\end{eqnarray}
\begin{equation}
\frac{d\eta ^{2}\left( Q\right) }{dQ}-\frac{d\zeta \left( r\right) }{dr}=2%
\frac{\zeta \left( r\right) }{r}+\frac{\eta ^{1}\left( U\right) }{U}.
\label{X2}
\end{equation}

In Eq. (\ref{X2}) the variables can be easily separated, leading to general
expressions for the functions $\zeta \left( r\right) $, $\eta ^{1}\left(
U\right) $ and $\eta ^{2}\left( Q\right) $ of the form
\begin{equation}
\zeta \left( r\right) =\frac{a}{r^{2}}+br,\eta ^{1}\left( U\right) =\left(
c-3b\right) U,\eta ^{2}\left( Q\right) =cQ+d,
\end{equation}
where $a,b,c,d$ are separation and integration constants, respectively.
Substituting in Eq. (\ref{X1}) gives the following consistency condition for
the coefficients $a,b,c,d$ and for the functions $\gamma \left( U\right) $
and $c_{s}$:
\begin{eqnarray}\label{Xf}
\frac{4a}{r^{3}}+b &=&\left( c-3b\right) U\left( \frac{\gamma ^{\prime }}{%
\gamma }-\frac{2c_{s}^{\prime }}{1+2c_{s}}\right) -\frac{a}{r^{4}}\frac{2GM+Q}{%
1-\frac{2GM}{r}-\frac{Q}{r}}+\frac{1}{r}\frac{\left( d-2bGM\right)
+\left( c-b\right) Q}{1-\frac{2GM}{r}-\frac{Q}{r}}+  \nonumber\\
&&\frac{a}{r^{2}}\frac{3\alpha \left( 1+2\gamma ^{-1}\right)
\left( 1+2\gamma \right) Ur^{2}+6}{\left( 1+2\gamma ^{-1}\right)
\left[ 2GM+Q+\alpha \left( 1+2\gamma \right) Ur^{3}\right]
+6r-6\left( 2GM+Q\right) }+  \nonumber\\
&&\frac{\left( 1+2\gamma ^{-1}\right) \left[ 3\alpha b\left(
1+2\gamma \right) Ur^{3}+cQ+d\right] +6br-6\left( cQ+d\right)
}{\left( 1+2\gamma ^{-1}\right) \left[ 2GM+Q+\alpha \left(
1+2\gamma \right) Ur^{3}\right]
+6r-6\left( 2GM+Q\right) }+  \nonumber \\
&&\left( c-3b\right) U\frac{-2\gamma ^{\prime }\gamma ^{-2}\left[
2GM+Q+\alpha \left( 1+2\gamma \right) Ur^{3}\right] +\alpha \left(
1+2\gamma ^{-1}\right)\left( 1+2\gamma +2\gamma ^{\prime }U\right)
r^{3}}{\left( 1+2\gamma ^{-1}\right) \left[ 2GM+Q+\alpha \left(
1+2\gamma \right) Ur^{3}\right] +6r-6\left( 2GM+Q\right) }.
\end{eqnarray}

Eq. (\ref{Xf}) is identically satisfied for $a=b=c=d=0$, corresponding to $%
{\bf X}=0$ (the identity transformation). In the second case, in order to
satisfy the identity (\ref{Xf}), we have to take as a first step $a=0$. Then we chose $%
d=2bGM$ and $c=b$. From the general structure of the identity
(\ref{Xf}) it follows that it cannot be identically satisfied
unless $\gamma ^{\prime }=0$
and $c_{s}^{\prime }=0$, implying $\gamma =P/U=$constant and $\gamma =c_{s}=$%
constant. Then in order that (\ref{Xf}) to be identically
satisfied we take for $b$ the value $b=1$ as the last
compatibility condition. Therefore we have obtained the following
theorem:

{\bf Theorem}. The static, spherically symmetric gravitational field
equations in vacuum on the brane are invariant with respect to the group of
the quasi-homologous transformations if and only if the dark pressure is
proportional to the dark radiation, $P=\gamma U$, $\gamma =$constant.

The infinitesimal operator generating the group of quasi-homologous
transformations on the static brane has the form
\begin{equation}
{\bf X}=r\frac{\partial }{\partial r}-2U\frac{\partial }{\partial
U}+\left( Q+2GM\right) \frac{\partial }{\partial Q}.
\end{equation}

The quantities $\left( Q+2GM\right) /r$ and $Ur^{2}$ (or any two
independent functions of them) are homologous invariants. Hence
the homology properties of the spherically symmetric vacuum
space-times on the brane are described by the following homology
theorem:

{\bf Homology Theorem.} If $U\left( r\right) $ is a solution of the static,
spherically symmetric gravitational field equations in vacuum on a brane
with the dark pressure proportional to the dark radiation, then so also is $%
C^{2}U\left( Cr\right) $, where $C$ is an arbitrary constant.

\section{Discussions and final remarks}

In the present paper we have considered some properties of the vacuum
exterior to compact astrophysical objects in the brane world model. The
system of field equations can be reduced to two ordinary differential
equations, in three unknowns, whose solution gives all the geometrical
properties of the space-time. The system of basic equations describing the
vacuum gravitational field equation on the brane is not uniquely determined
and its solution depends on the functional relation between two unknown
functions, the dark pressure $P$ and the dark radiation $U$. The symmetry
properties of the vacuum brane space-times can uniquely fix the functional
relation between these two free parameters of the model. The requirement
that the vacuum on the brane admits a one-parameter group of conformal
motions or a group of homologous transformations uniquely fixes the
functional dependence of the free parameters $P$ and $U$. The relation
between the dark pressure and the dark radiation for the vacuum on the brane
admitting a one-parameter group of conformal motions is of the form $%
2P+U=\left( B/\alpha \right) r^{-4}$. On the other hand the invariance of
the field equations with respect to the Lie group of homologous
transformations requires a linear proportionality relation between $P$ and $%
U $, $P=\gamma U$. Once the relation between $P$ and $U$ is known, the
general solution of the vacuum field equations can be found perturbatively.

The Schwarzschild solution is no longer the unique vacuum solution of the
gravitational field equations. Moreover, most of the general solutions we
have found are not asymptotically flat and consequently they are of
cosmological nature.

In order to obtain a manifestly coordinate invariant characterization of
certain geometrical properties of geometries, like for example curvature
singularities, Petrov type of the Weyl tensor etc. the scalar invariants of
the Riemann tensor have been extensively used. Two scalars, which have been
considered in the physical literature, are the Kretschmann scalars, $%
RiemSq\equiv R_{ijkl}R^{ijkl}$ and $RicciSq\equiv R_{ij}R^{ij}$, where $%
R_{ijkl}$ is the Riemann curvature tensor.

For space-times which are the product of two 2-dimensional spaces, one
Lorentzian and one Riemannian, subject to a separability condition on the
function which couples the 2-spaces, it has been suggested \cite{Sa98} that
the set
\begin{equation}\label{set}
C=\left\{ R,r_{1},r_{2},w_{2}\right\} ,
\end{equation}
form an independent set of scalar polynomial invariants, satisfying the
number of degrees of freedom in the curvature. In Eq. (\ref{set}) $%
R=g^{il}g^{jk}R_{ijkl}$ is the Ricci scalar and the quantities $r_{1}$, $%
r_{2}$ and $w_{2}$ are defined according to \cite{Za01}
\begin{equation}
r_{1}=\phi _{AB\dot{A}\dot{B}}\phi ^{AB\dot{A}\dot{B}}=\frac{1}{4}%
S_{a}^{b}S_{b}^{a},
\end{equation}
\begin{equation}
r_{2}=\phi _{AB\dot{A}\dot{B}}\phi _{C\dot{C}}^{B\dot{B}}\phi ^{CA\dot{C}%
\dot{A}}=-\frac{1}{8}S_{a}^{b}S_{b}^{c}S_{c}^{a},
\end{equation}
\begin{equation}
w_{2}=\Psi _{ABCD}\Psi _{EF}^{CD}\Psi ^{EFAB}=-\frac{1}{8}\bar{C}_{abcd}\bar{%
C}_{ef}^{cd}\bar{C}^{efab},
\end{equation}
where $S_{a}^{b}=R_{a}^{b}-\frac{1}{4}R\delta _{a}^{b}$ is the trace-free
Ricci tensor, $\phi _{AB\dot{A}\dot{B}}$ denotes the spinor equivalent of $%
S_{ab}$, $\Psi _{ABCD}$ denotes the spinor equivalent of the Weyl tensor $%
C_{abcd}$ and $\bar{C}_{abcd}$ denotes the complex conjugate of the
self-dual Weyl tensor, $C_{abcd}^{+}=\frac{1}{2}\left( C_{abcd}-i\ast
C_{abcd}\right) $.

In terms of the ''electric'' $E_{ac}=C_{abcd}u^{b}u^{d}$ and ''magnetic'' $%
H_{ac}=C_{abcd}^{\ast }u^{b}u^{d}$ parts of the Weyl tensor, where $u^{a}$
is a timelike unit vector and $C_{abcd}^{\ast }=\frac{1}{2}\eta
_{abef}C_{cd}^{ef}$ is the dual tensor, the invariant $w_{2}$ is given by
\cite{Sa98}
\begin{equation}
w_{2}=\frac{1}{32}\left(
3E_{b}^{a}H_{c}^{b}H_{a}^{c}-E_{b}^{a}E_{c}^{b}E_{a}^{c}\right) +\frac{i}{32}%
\left( H_{b}^{a}H_{c}^{b}H_{a}^{c}-3E_{b}^{a}E_{c}^{b}H_{a}^{c}\right) .
\end{equation}

The values of the invariant set $\left\{ R,r_{1},r_{2},w_{2}\right\} $ for
some static spherically symmetric vacuum brane solutions are presented in
the Appendix.

The corrections to the Newtonian potential on the brane have been considered
by using perturbative expansions in the static weak-field regime. The
leading order correction to the Newtonian potential on the brane is given by
$\Phi =\left( GM/r\right) \left( 1+2l^{2}/3r^{2}\right) $ \cite{RS99a},
where $l$ is the curvature scale of the five-dimensional anti de Sitter
spacetime (AdS$_{5}$), or it can also involve a logarithmic factor \cite
{Ga00}. This type of weak-field behavior cannot be recovered in the classes
of solutions we have considered in the present paper. However, this could be
possible for models involving a more precise knowledge of the general
behavior of the dark radiation and dark pressure terms.

\section*{Appendix}

In this Appendix we present the values of the Kretschmann scalars $%
RiemSq\equiv R_{ijkl}R^{ijkl}$ and $RicciSq\equiv R_{ij}R^{ij}$and some
values of the independent set of the scalar polynomial invariants $\left\{
R,r_{1},r_{2},w_{2}\right\} $ for the exact static spherically symmetric
vacuum brane geometries discussed in the paper.

a)
\begin{equation}
e^{\nu }=e^{-\lambda }=1-\frac{2GM+Q_{0}}{r}-\frac{3\alpha P_{0}}{2r^{2}},
\end{equation}
\begin{equation}
RicciSq=\frac{9\alpha ^{2}P_{0}}{r^{8}},RiemSq=\frac{6\left[ 21\alpha
^{2}P_{0}^{2}+12\alpha P_{0}\left( 2GM+Q_{0}\right) r+2\left(
2GM+Q_{0}\right) ^{2}r^{2}\right] }{r^{8}},
\end{equation}
\begin{equation}
R=0,r_{1}=\frac{9\alpha ^{2}P_{0}}{4r^{8}},r_{2}=0,
\end{equation}
\begin{equation}
Re\left(w_{2}\right)=\frac{3\left[ 27\alpha ^{3}P_{0}^{3}+27\alpha
^{2}P_{0}^{2}\left( 2GM+Q_{0}\right) r+9\alpha P_{0}\left(
2GM+Q_{0}\right) ^{2}r^{2}+\left( 2GM+Q_{0}\right)
^{3}r^{3}\right] }{4r^{12}}, Im\left(w_{2}\right)=0.
\end{equation}

b)
\begin{equation}
e^{\nu }=C_{2}\left( 1-\frac{2GM}{r}\right) \left[ C_{1}-\frac{\alpha }{GM}%
\left( 1-\frac{2GM}{r}\right) ^{-1/2}\right] ^{2},e^{-\lambda }=1-\frac{2GM}{%
r},
\end{equation}
\begin{equation}
RicciSq=\frac{6\alpha ^{2}}{\left( \frac{\alpha }{GM}-C_{1}\sqrt{1-\frac{2GM%
}{r}}\right) ^{2}r^{6}},
\end{equation}
\begin{equation}
RiemSq=\frac{24\left[ -4C_{1}^{2}G^{3}M^{3}+\left( \alpha
^{2}+2C_{1}^{2}G^{2}M^{2}-2\alpha C_{1}GM\sqrt{1-\frac{2GM}{r}}\right) r%
\right] }{\left( \frac{\alpha }{GM}-C_{1}\sqrt{1-\frac{2GM}{r}}\right)
^{2}r^{7}},
\end{equation}
\begin{equation}
R=0,r_{1}=\frac{3\alpha ^{2}}{2\left( \frac{\alpha }{GM}-C_{1}\sqrt{1-\frac{%
2GM}{r}}\right) ^{2}r^{6}},r_{2}=-\frac{3\alpha ^{3}}{4\left( \frac{\alpha }{%
GM}-C_{1}\sqrt{1-\frac{2GM}{r}}\right) ^{3}r^{9}},
\end{equation}
\begin{eqnarray}
Re\left( w_{2}\right)  &=&\frac{3}{4}\frac{9\alpha
C_{1}G^{2}M^{2}\left( 2GM-r\right) \left[ \left( \frac{\alpha }{GM}\right)
^{4}r^{2}+7\left( \frac{\alpha }{GM}\right) ^{2}C_{1}^{2}r\left(
r-2GM\right) +4C_{1}^{4}\left( 2GM-r\right) ^{2}\right] }{\left( C_{1}\sqrt{%
1-\frac{2GM}{r}}-\frac{\alpha }{GM}\right) ^{6}\sqrt{1-\frac{2GM}{r}}r^{12}}-
\nonumber\\
&&\frac{3}{4}\frac{G^{3}M^{3}\left[ 2C_{1}^{2}\left( 2GM-r\right) -\left(
\frac{\alpha }{GM}\right) ^{2}r\right] \left[ \left( \frac{\alpha }{GM}%
\right) ^{4}r^{2}-31\left( \frac{\alpha }{GM}\right) ^{2}C_{1}^{2}r\left(
2GM-r\right) +4C_{1}^{2}\left( 2GM-r\right) ^{2}\right] }{\left( C_{1}\sqrt{%
1-\frac{2GM}{r}}-\frac{\alpha }{GM}\right) ^{6}r^{12}},
\end{eqnarray}
\begin{equation}
Im\left(w_{2}\right)=0.
\end{equation}

c)
\begin{equation}
e^{\nu }=A^{2}r^{2},e^{-\lambda }=\frac{1}{3}\left( 1+\frac{B}{r^{2}}\right)
,
\end{equation}
\begin{equation}
RicciSq=\frac{2\left( 2B^{2}+r^{4}\right) }{3r^{8}},RiemSq=\frac{8\left(
B^{2}+r^{4}\right) }{3r^{8}},
\end{equation}
\begin{equation}
R=0,r_{1}=\frac{2B^{2}+r^{4}}{6r^{8}},r_{2}=\frac{4B^{3}-3Br^{4}-r^{6}}{%
36r^{12}},
\end{equation}
\begin{equation}
Re\left(w_{2}\right)=\frac{1}{36r^{6}}, Im\left(w_{2}\right)=0.
\end{equation}


\begin{references}

\bibitem{RS99a}  L. Randall and R. Sundrum, {\it Phys. Rev. Lett. {\bf 83}},
3370 (1999); L. Randall and R. Sundrum, {\it Phys. Rev. Lett {\bf 83}}, 4690
(1999).

\bibitem{HW96}  P. Horava and E. Witten, {\it Nucl. Phys. {\bf B460}}, 506
(1996).

\bibitem{Ma01}  R. Maartens, {\tt gr-qc/0101059}.

\bibitem{SMS00}  T. Shiromizu, K. Maeda and M. Sasaki, {\it Phys. Rev. {\bf %
D62}}, 024012 (2000).

\bibitem{SSM00}  M. Sasaki, T. Shiromizu and K. Maeda, {\it Phys. Rev. {\bf %
D62}}, 024008 (2000).

\bibitem{all}  K. Maeda and D. Wands, {\it Phys. Rev. {\bf D62}}, 124009
(2000); R. Maartens, {\it Phys. Rev. {\bf D62}}, 084023 (2000); A.
Campos and C. F. Sopuerta, {\it Phys. Rev. {\bf D63}}, 104012
(2001); A. Campos and C. F. Sopuerta, {\it Phys. Rev. {\bf D64}},
104011 (2001); C.-M. Chen, T. Harko and M. K. Mak, {\it Phys. Rev.
{\bf D64, }} 044013 (2001); D. Langlois, {\it Phys. Rev. Lett.
{\bf 86}}, 2212 (2001); C.-M. Chen, T. Harko and M. K. Mak, {\it
Phys. Rev. {\bf D64}}, 124017 (2001);  L. Anchordoqui, J.
Edelstein, C. Nunez, S. P. Bergliaffa, M. Schvellinger, M. Trobo
and F. Zyserman, {\it Phys. Rev.} {\bf D64}, 084027 (2001); A.
Coley, {\it Phys. Rev. {\bf D66, }} 023512 (2002); A. Coley, {\it
Class. Quantum Grav. {\bf 19 }}, L45 (2002); J. D. Barrow and R.
Maartens, {\it Phys. Lett. {\bf B532}}, 153 (2002); H. Kudoh and
T. Tanaka, {\it Phys. Rev. {\bf D65}}, 104034 (2002); H. A.
Bridgman, K. A. Malik and D. Wands, {\it Phys. Rev. {\bf D65}},
043502 (2002); C.-M. Chen, T. Harko, W. F. Kao and M. K. Mak, {\it Nucl. Phys. {\bf %
B64, }}159 (2002); M. Szydowski, M. P. Dabrowski and A. Krawiec, {\it Phys. Rev.} {\bf D66}, 064003 (2002);
 M. K. Mak and T. Harko, {\it Class. Quantum Grav. {\bf 20}%
}, 407 (2003);  I. Brevik and A. Hallanger, gr-qc/0308058; C.-M.
Chen, T. Harko, W. F. Kao and M. K. Mak, {\it JCAP} {\bf 0311},
005 (2003); I. Brevik, K. Børkje and J. P. Morten, gr-qc/0310103.

\bibitem{all1}  N. Dadhich and S. G. Ghosh, {\it Phys. Lett.} {\bf B518, }1
(2001); M. G. Santos, F. Vernizzi and P. G. Ferreira, {\it Phys. Rev.} {\bf %
D64}, 063506 (2001); M. Bruni, C. Germani and R. Maartens, {\it Phys. Rev.
Lett.} {\bf 87}, 231302 (2001); H.-C. Kim, S.-H. Moon and J. H. Yee, {\it %
JHEP} {\bf 0202}, 046 (2002); M Govender and N. Dadhich, {\it
Phys. Lett.} {\bf B538}, 233 (2002); T. Wiseman, {\it Class.
Quant. Grav.} {\bf 19}, 3083 (2002); R. Neves and C. Vaz, {\it
Phys. Rev. {\bf D66}},124002 (2002);  L. A. Anchordoqui, H.
Goldberg and A. D. Shapere, {\it Phys. Rev.} {\bf  D66}, 024033
(2002); H. Kudoh, T. Tanaka and T. Nakamura, gr-qc/0301089.

\bibitem{Da00}  N. Dadhich, R. Maartens, P. Papadopoulos and V. Rezania,
{\it Phys. Lett. {\bf B487}}, 1 (2000).

\bibitem{GeMa01}  C. Germani and R. Maartens, {\it Phys. Rev. {\bf D64, }}
124010 (2001).

\bibitem{Ca02}  R. Casadio, A. Fabbri and L. Mazzacurati, {\it Phys. Rev.}
{\bf D65}, 084040 (2002).

\bibitem{Da03}  S. Shankaranarayanan and N. Dadhich, gr-qc/0306111 (2003).

\bibitem{Vi03}  M. Visser and D. L. Wiltshire, {\it Phys. Rev.} {\bf D67},
104004 (2003).

\bibitem{Ca03}  R. Casadio and L. Mazzacurati, {\it Mod. Phys. Lett.} {\bf %
A18}, 651 (2003).

\bibitem{Cha59}  S. Chandrasekhar, {\it An introduction to the study of
stellar structure}, New York, Dover Publications (1959).

\bibitem{Makm}  M. K. Mak, H. W. Chan and T. Harko, {\it Comp. Math.
Applications }{\bf 41}, 1395 (2001); M. K. Mak and T. Harko, {\it
Comp. Math. Applications }{\bf 43}, 91 (2002); T. Harko and M. K.
Mak, {\it Comp. Math. Applications }{\bf 46}, 849 (2003).

\bibitem{He}  L. Herrera L. and J. Ponce de Leon, {\it J. Math. Phys.} {\bf %
26}, 2303 (1985); L. Herrera L. and J. Ponce de Leon, {\it J.
Math. Phys.} {\bf 26}, 2018 (1985); L. Herrera L. and J. Ponce de
Leon, {\it J. Math. Phys.} {\bf 26}, 778 (1985); M. K. Mak and T.
Harko, to appear in {\it Int. Journal Mod. Phys.} {\bf D}, {\bf \
}{\it gr-qc/0309069} (2003).

\bibitem{Ol93}  P. T. Olver, {\it Applications of Lie groups to differential
equations}, New York, Springer-Verlag (1993); B. J. Cantwell, {\it %
Introduction to symmetry analysis}, Cambridge, Cambridge University Press
(2002).

\bibitem{Co77}  C. B. Collins, {\it J. Math. Phys.} {\bf 18}, 1374 (1977).

\bibitem{Sa98}  K. Santusuosso, D. Pollney, N. Pelavas, P. Musgrave and K.
Lake, {\it Comp. Phys. Comm.} {\bf 115}, 381 (1998).

\bibitem{Za01}  E. Zakhary and J. Carminati, {\it J. Math. Phys.} {\bf 42}
1474 (2001); J. Carminati, E. Zakhary and R. G. McLenaghan, {\it J. Math.
Phys.} {\bf 43} 492 (2002).

\bibitem{Ga00}  J. Garriga and T. Tanaka, {\it Phys. Rev. Lett}.{\bf 84},
2778 (2000); S. Nojiri and S. D. Odintsov, {\it Phys. Lett.} {\bf B548},215
(2002); M. Ito, {\it Phys. Lett.} {\bf B528}, 269 (2002); M. Ito, {\it Phys.
Lett.} {\bf B554}, 180 (2003); D. K. Park, {\it Phys. Lett.} {\ bf B562},
316 (2003); D. K. Park and S. Tamaryan, {\it Phys. Lett.} {\bf B554}, 92
(2003).

\end{references}
\end{document}